\documentclass[12pt,twoside]{article}

\usepackage{epsf} 

\def\PRL{{\em Phys.~Rev.~Lett.}}
\def\PRD{{\em Phys.~Rev.}~D}

\def\be{\begin{equation}}
\def\ee{\end{equation}}
\def\bea{\begin{eqnarray}}
\def\eea{\end{eqnarray}}

\newcommand{ \centeron }[2]{{\setbox0=\hbox{#1}\setbox1=\hbox{#2}\ifdim
                             \wd1>\wd0\kern.5\wd1\kern-.5\wd0\fi \copy0
                             \kern-.5\wd0\kern-.5\wd1\copy1\ifdim\wd0>\wd1
                             \kern.5\wd0\kern-.5\wd1\fi}}
\newcommand{ \ltap }{\;\centeron{\raise.35ex\hbox{$<$}}
                     {\lower.65ex\hbox{$\sim$}}\;}
\newcommand{\sss}{\scriptscriptstyle}
\newcommand{\epem}{e^{\sss +}e^{\sss -}}   
\def\G{\tilde{G}}
\newcommand{\NI }{\tilde{N}_1}
\newcommand{\NII}{\tilde{N}_2}
\newcommand{\CI }{\tilde{C}_1}
\newcommand{\eL }{{\tilde e}_{\scriptscriptstyle L}}
\newcommand{\eR }{ {\tilde e}_{\scriptscriptstyle R}}
\newcommand{\tauu}{{\tilde \tau}_1}
\newcommand{\taud}{{\tilde \tau}_2}
\newcommand{\ra}{\rightarrow}

\begin{document}
\thispagestyle{plain} 

\centerline{hep-ph/9911360 \hfill CERN-TH/99-348}
\vspace*{0.1cm}
\hrule\hfill 

\begin{center}
{\Large \bf 
Extracting GMSB Parameters at a Linear Collider} 
\footnote{Work supported also by {\it Deutsches 
Elektronen-Synchrotron} DESY, Hamburg, Germany} 
\end{center}

\begin{center}
{\large {\bf Sandro~Ambrosanio} \\
        CERN -- {\it Theory Division}, \\
     CH-1211 Geneva 23, Switzerland \\
     e-mail: {\sl ambros@mail.cern.ch}} \\
~\\
{\large and} \\
~\\ 
{\large {\bf Grahame A.~Blair} \\ 
      {\it Royal Holloway and Bedford New College}, \\
      University of London, Egham Hill, Egham, \\
      Surrey TW20 0EX, U.K. \\
      e-mail: {\sl g.blair@rhbnc.ac.uk}} \\
\end{center}

\begin{center} 
{\bf Abstract} \\
\end{center}
{\small 
Assuming gauge-mediated supersymmetry (SUSY) breaking (GMSB),
we simulate precision measurements of fundamental
parameters at a 500 GeV $\epem$ linear collider (LC) in the scenario where a 
neutralino is the next-to-lightest supersymmetric particle (NLSP). 
Information on the SUSY breaking and the messenger sectors of the 
theory is extracted from realistic fits to the measured mass spectrum 
of the Minimal SUSY Model (MSSM) particles and the NLSP lifetime.}

\centerline{\hfill \parbox{0.4\textwidth}{\hrule\hfill} \hfill}
\begin{center}
{\sl Contribution to the Workshops: \\
2$^{\rm nd}$ ECFA/DESY Study on Physics and Detectors for a 
Linear $\epem$ Collider \\
Lund, Frascati, Oxford, Obernai -- June 1998 to October 1999} 
\end{center} 
\centerline{\hfill \parbox{0.4\textwidth}{\hrule\hfill} \hfill}
  
\noindent
Supersymmetry must be broken if it is to describe nature, and
GMSB~\cite{GRreport} is one attractive way to realize this, 
also providing natural suppression of the SUSY contributions 
to flavour-changing neutral currents at low energies. 
In GMSB models, the gravitino $\G$ is the LSP with mass given by 
$m_{\G} = \frac{F}{\sqrt{3}M'_P} \simeq 
2.37 \left(\frac{\sqrt{F}}{100 \; {\rm TeV}}\right)^2 \; {\rm eV}$,  
where $\sqrt{F}$ is the fundamental SUSY breaking scale. 
The GMSB phenomenology is characterised by decays of the NLSP to its 
Standard Model partner and the $\G$ with a non-negligible or even 
macroscopic lifetime. 
In the simplest GMSB realizations, depending on the parameters
$M_{\rm mess}$, $N_{\rm mess}$, $\Lambda$, $\tan\beta$, sign($\mu$)
defining the model, the NLSP can be either the lightest neutralino 
$\NI$ or the light stau $\tauu$. For this study~\cite{ourpaper},
we generated several thousand GMSB models following the standard 
phenomenological approach~\cite{AKM} and focused on the neutralino
NLSP scenario, for which we selected several representative points
for simulation. Our aim was to explore the potential of a LC
in extracting the fundamental model parameters. 

Firstly, we investigated the sensitivity in determining the
GMSB parameters at the messenger and electroweak scales from the knowledge 
of the sparticle masses that could be obtained from threshold-scanning 
techniques. We used a sample model with \\
$M_{\rm mess} = $ 161 TeV; $N_{\rm mess} = 1$; $\Lambda = 76$ TeV; 
$\tan\beta = 3.5$; $\mu > 0$, producing a rather light sparticle 
spectrum, and assumed a total of 200 fb$^{-1}$ collected between 200 
and 500 GeV c.o.m. energies at a LC. By just considering the shape of 
the total cross sections for several kinematically allowed SUSY production 
processes as functions of $\sqrt{s}$ close to the thresholds, we inferred
the following approximate precisions for the sparticle masses: 

\bea 
\Delta(m_{\NI})   \sim 0.2 \; {\rm GeV}; & 
\Delta(m_{\NII})  \sim 0.8 \; {\rm GeV}; & 
\Delta(m_{\CI})   \sim 0.1 \; {\rm GeV}; \nonumber \\
\Delta(m_{\eL})   \sim 0.2 \; {\rm GeV}; &
\Delta(m_{\eR})   \sim 0.2 \; {\rm GeV}; &
\Delta(m_{\mu_R}) \sim 0.8 \; {\rm GeV}; \label{eq:errmass} \\
\Delta(m_{\tauu}) \sim 0.8 \; {\rm GeV}; &
\Delta(m_{\taud}) \sim 2.0 \; {\rm GeV}; &
\Delta(m_{h^0})   \sim 0.1 \; {\rm GeV}. \nonumber
\eea

By performing a fit to minimise a $\chi^2$ based on these errors,
with the true (model-dependent) values as the central ones in the fit, 
we obtained an estimate of the precisions on the underlying parameters, 
as shown in Tab.~\ref{tab:GMSBfit}.  
We checked that these precisions are typical for the class of models we
considered.

\begin{table}
\renewcommand{\arraystretch}{1.2}
\begin{center}
\begin{tabular}{|c||c|}                          \hline
Parameter      & Fitted value                 \\ \hline \hline  
$M_{\rm mess}$ & ($161    \pm 2$)      TeV    \\ \hline
$\Lambda$      & ($76.01  \pm 0.08$)   TeV    \\ \hline
$N_{\rm mess}$ & $0.9994 \pm 0.0009$          \\ \hline
$\tan\beta$    & $3.50   \pm 0.03$            \\ \hline
\end{tabular}
\end{center}
\caption{\sl
Results of fits to the parameters of the GMSB model described
in the text, starting from a  possible set of light sparticle masses 
measurements from threshold scans. A 200 fb$^{-1}$ run at the 
LC is assumed.}
\label{tab:GMSBfit}
\end{table}

Then, we considered $\NI$ lifetime measurements in the whole allowed 
$c\tau_{\NI}$ range, performing event simulation in detail for our 
set of representative GMSB models. 
Indeed, since the $\NI$ lifetime is related to $\sqrt{F}$ by
\be
c \tau_{\NI} = \frac{16\pi}{{\cal B}} \frac{\sqrt{F}^4}{m_{\NI}^5}
\simeq \frac{1}{100 {\cal B}} 
\left(\frac{\sqrt{F}}{100 \; {\rm TeV}}\right)^4 
\left(\frac{m_{\NI}} {100 \; {\rm GeV}}\right)^{-5},
\label{eq:NLSPtau}
\ee 
the GMSB framework provides an opportunity to extract information on the 
SUSY breaking sector of the theory from collider experiments that is not 
available, e.g., in supergravity-inspired models. 

\begin{figure}
\centerline{
\epsfxsize=3.0in
\epsffile{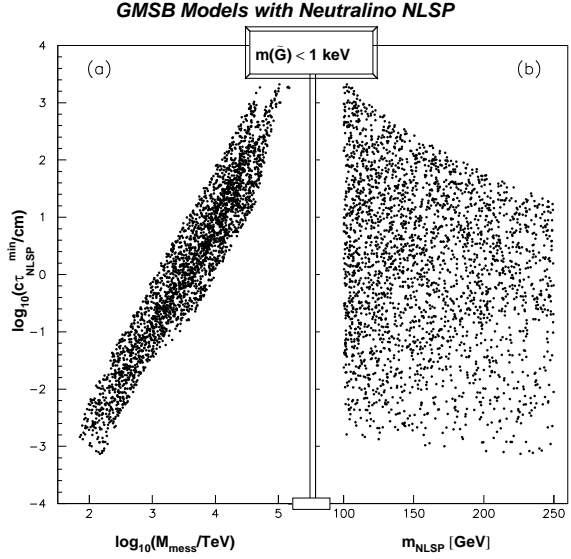} 
}
\caption{\sl Scatter plot of the neutralino NLSP lifetime as a function of 
the messenger scale $M_{\rm mess}$ (a) and $m_{\NI}$ (b). 
For each set of GMSB model input parameters, we plot the lower limit on 
$c\tau_{\NI}$, corresponding to 
$\sqrt{F} \simeq \sqrt{F}_{\rm mess} = \sqrt{\Lambda M_{\rm mess}}$. 
We use only models that fulfil the limit 
$m_{\G} \ltap 1 \; {\rm keV} \Rightarrow 
\sqrt{F}_{\rm mess} \ltap \sqrt{F} \ltap 2000$ TeV 
suggested by simple cosmology. 
}
\label{fig:one}
\end{figure}

Typical neutralino lifetimes for our models range from microns to 
tens of metres. While the lower bound on $c\tau_{\rm \NI}$ comes from 
requiring perturbativity up to the grand unification scale~\cite{AKM}, the 
upper bound is only valid if the $\G$ mass is restricted to be lighter than 
about 1 keV, as suggested by some cosmological arguments \cite{Cosmo} 
(cfr. Fig.~\ref{fig:one}).

\begin{figure}
\centerline{
\epsfxsize=3.0in
\epsffile{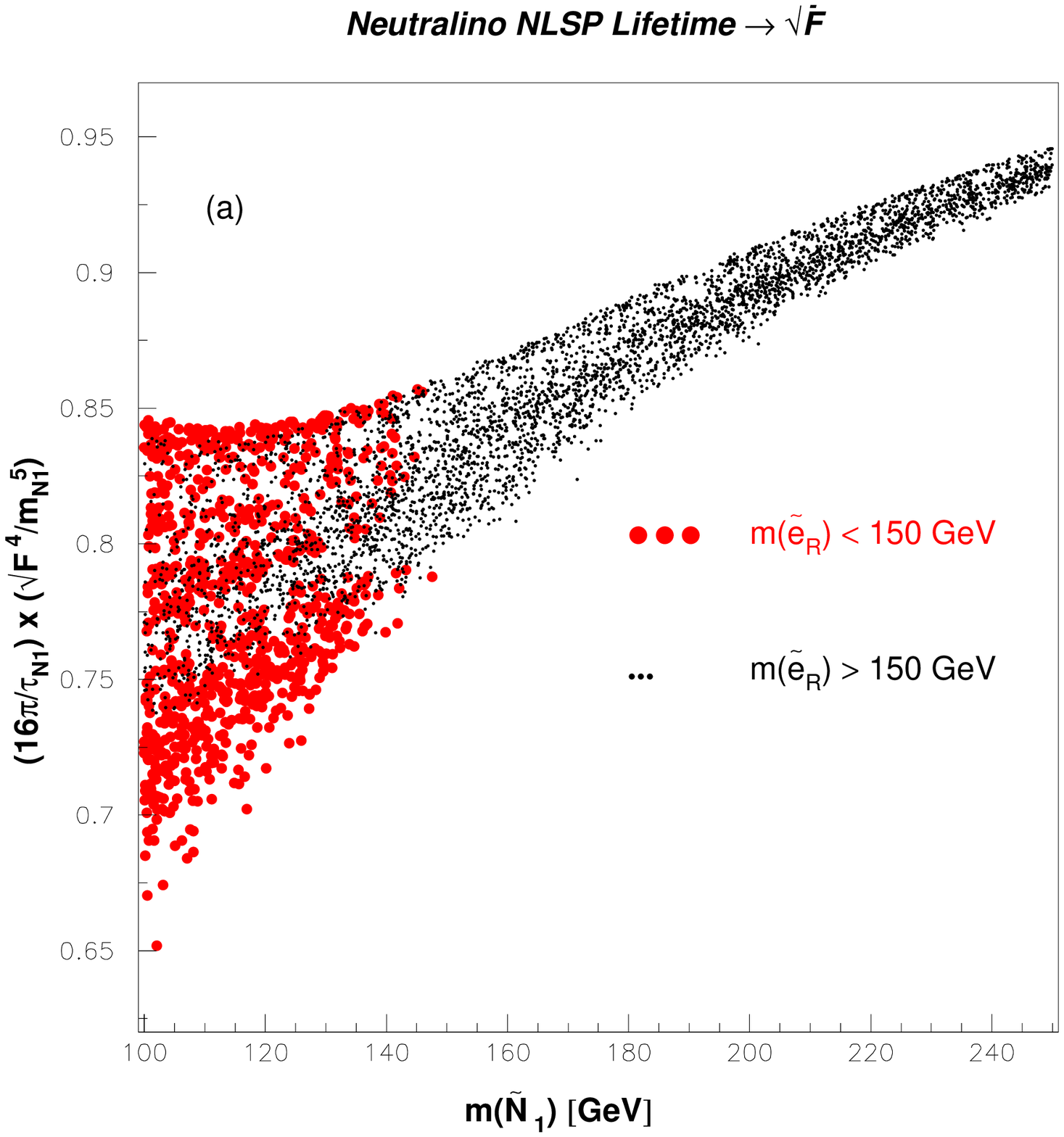} 
\hfill
\epsfxsize=3.0in
\epsffile{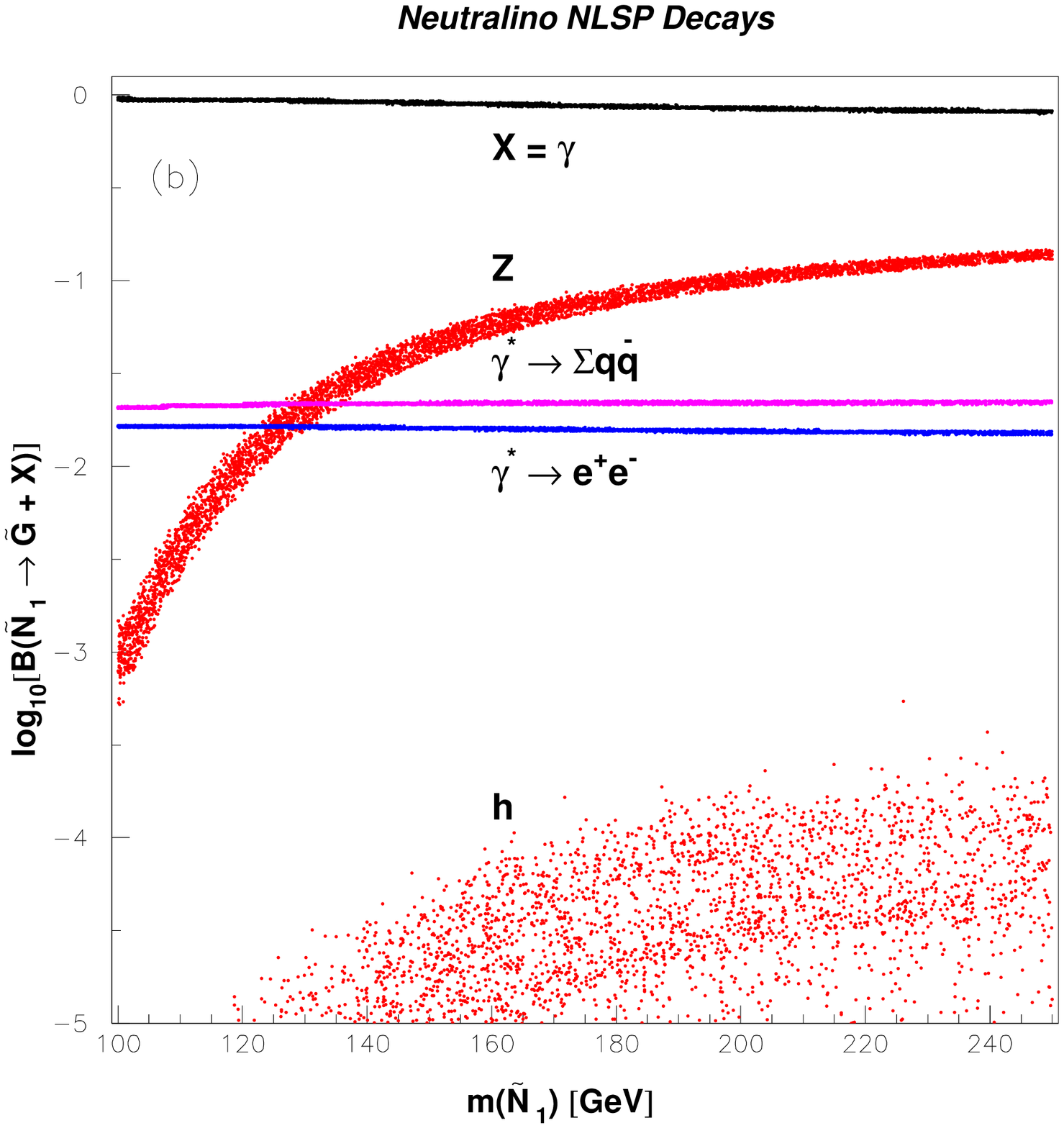} 
}
\caption{\sl (a) Scatter plot showing the relation between the neutralino NLSP
lifetime and the fundamental scale of SUSY breaking $\sqrt{F}$, i.e. the 
factor ${\cal B}$ in Eq.(\ref{eq:NLSPtau}), as a function of the 
neutralino mass in GMSB models of interest for the LC 
(100 $\ltap m_{\NI} \ltap 250$ GeV).  
Big grey dots in represent models with a light $R$-selectron (102--150 GeV), 
small black dots are for the heavier selectron 
case (150--430 GeV). 
(b) Scatter plot for the BR's of various $\NI$ decay channels as 
a function of the $\NI$ mass. Dots in different grey scale (colours) refer 
to the decays $\NI \ra \gamma\G$, $\NI\ra Z \G$ (including off-shell effects), 
and to hadrons or $\epem$ plus gravitino via virtual photon, as labelled. 
For reference, we also report results for the 2-body $\NI\to h^0\G$ decay 
in the on-shell approximation, whose BR is always negligible.} 
\label{fig:two}
\end{figure}

 For given $\NI$ mass and lifetime, the residual theoretical uncertainty 
on determining $\sqrt{F}$ is due to the factor of order unity ${\cal B}$ 
in Eq.~(\ref{eq:NLSPtau}), whose variation is quite limited in GMSB 
models (cfr. Fig.~\ref{fig:two}a). 

In addition to the dominant $\NI\ra\gamma\G$ decay, it was fundamental 
to our analysis to take the $\NI\ra\G f \bar{f}$ decays into account, 
in order to use the tracking detectors for measurements of shorter 
neutralino lifetimes. 
We performed a complete study of these channels and found that in most
cases of interest for our study the total width is given approximately by 
\be
\Gamma(\NI\ra f\bar{f}\G) \simeq \Gamma(\NI\ra\gamma\G) 
\frac{\alpha_{\rm em}}{3\pi} N_f^c Q_f^2 
\left[2 \; {\rm ln}\frac{m_{\NI}}{m_f} - \frac{15}{4} \right]
+ \Gamma(\NI\ra Z\G)B(Z\ra f\bar{f}) \; , 
\label{eq:Steve-for}
\ee
where the expressions for the widths of the 2-body $\NI$ decays 
are well-known~\cite{AKKMM}. In Fig.~\ref{fig:two}b, the branching ratio
(BR) of the $\NI\ra \gamma\G$ decay is compared to those of $\NI\ra Z\G$ 
and $\NI\ra h^0 \G$ (in the on-shell approximation) and those of the main 
$\NI\ra f \bar{f}\G$ channels (including virtual-photon exchange 
contributions only). 

 To generate GMSB events, we modified {\tt SUSYGEN 2.2/03}~\cite{susygen}, 
to take the 3-body neutralino decays into account as follows. 
We implemented in {\tt CompHEP 3.3.18}~\cite{CompHEP} a home-made lagrangian 
including the relevant gravitino interaction vertices in a suitable 
approximation. Then, we studied the kinematical distributions 
of the $\NI\to f \bar{f}\G$ channels and passed the results to the event 
generator numerically. 
For each sample GMSB model, we considered in most cases a LC run at a 
c.o.m. energy such that the only SUSY production process open is 
NLSP pair production $\epem\ra\NI\NI$, followed by $\NI$ decays through  
all possible channels. For more challenging models where the light SUSY 
thresholds are close to each other, we simulated also events from
$R$-slepton pair production and used some selection to isolate the 
$\NI\NI$ events, for which the $\NI$ production energy is fixed by the 
beam energy (we also took into account initial-state radiation as well
as beamstrahlung effects), allowing a cleaner $c\tau_{\NI}$ measurement. 

 The primary vertex of the events was first smeared according to the assumed 
beamspot size of 5 nm in $y$, 500 nm in $x$ and  400 $\mu$m in $z$ and then 
the events were passed through a full {\tt GEANT 3.21}~\cite{Geant321} 
simulation of the detector as described in the ECFA/DESY CDR~\cite{CDR}.
The tracking detector components essential to our analysis included a 
5-layer vertex detector with a point precision of 3.5 $\mu$m in $r\phi$ 
and $z$, a TPC possessing 118 padrows with point resolution of 160 $\mu$m 
in $r\phi$ and 0.1 cm in $z$.
In addition, we assumed an electromagnetic calorimeter with energy resolution
given by ($10.3/\sqrt{E} + 0.6$)\%, angular pointing resolution of
$50/\sqrt{E}$ mrad and timing resolution of $2/\sqrt{E} $ ns.  
The dimensions of the whole calorimeter (electromagnetic and hadronic) 
were 172 cm $< r < 210$ cm and 280 cm $ <|z| < 330$ cm.  

 The probability for a single neutralino produced with energy $E_{\NI}$ 
to decay before travelling a distance $\lambda$ is given by 
$P(\lambda) = 1 - {\rm exp}(-\lambda/L)$, where $L = c \tau_{\NI} 
(\beta\gamma)_{\NI}$ is the  $\NI$ ``average'' decay length and 
$(\beta\gamma)_{\NI} = (E_{\NI}^2/m_{\NI}^2 - 1)^{1/2}$. 

\begin{figure}
\centerline{
\epsfxsize=3.0in
\epsffile{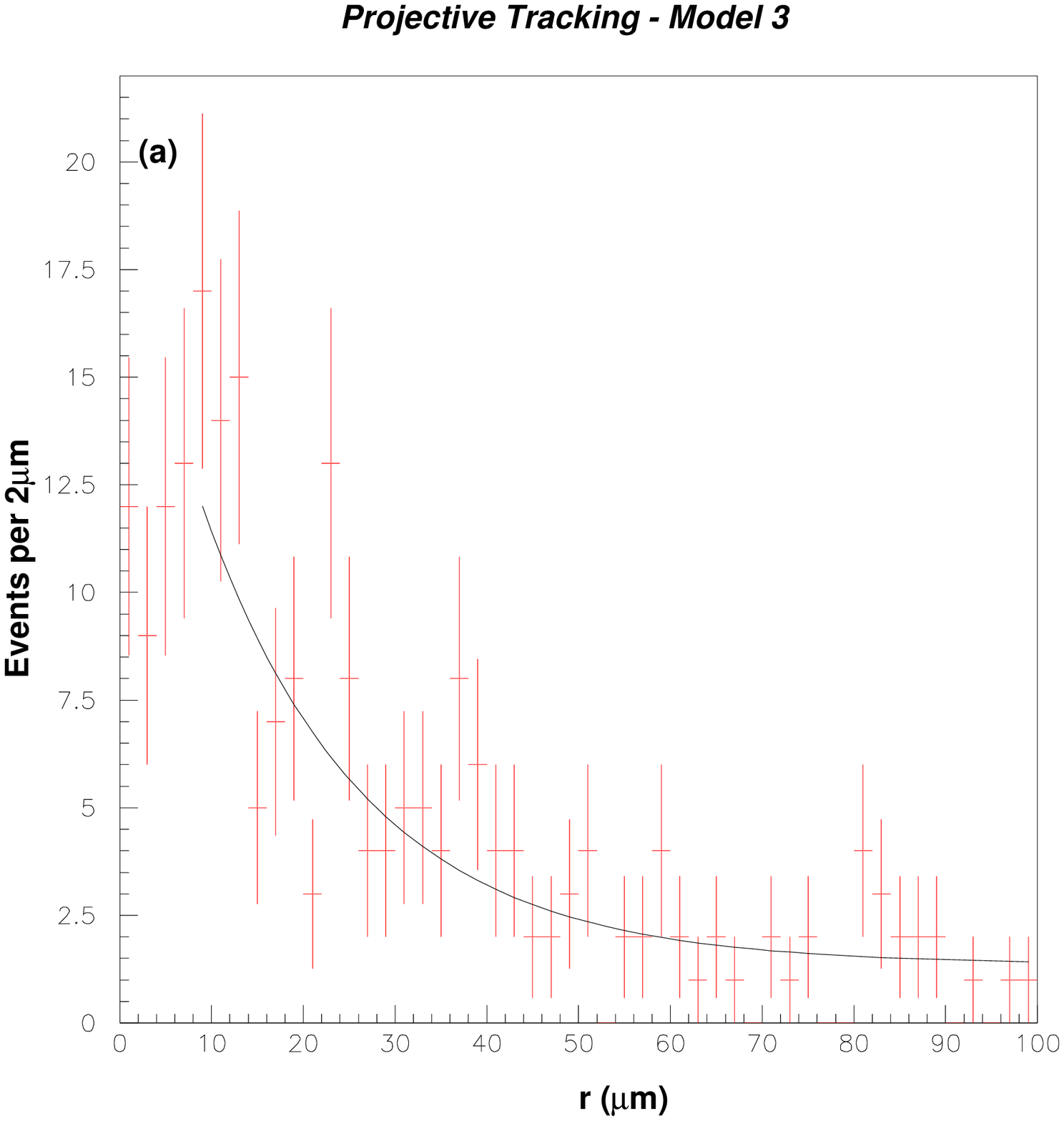} 
\hfill
\epsfxsize=3.0in
\epsffile{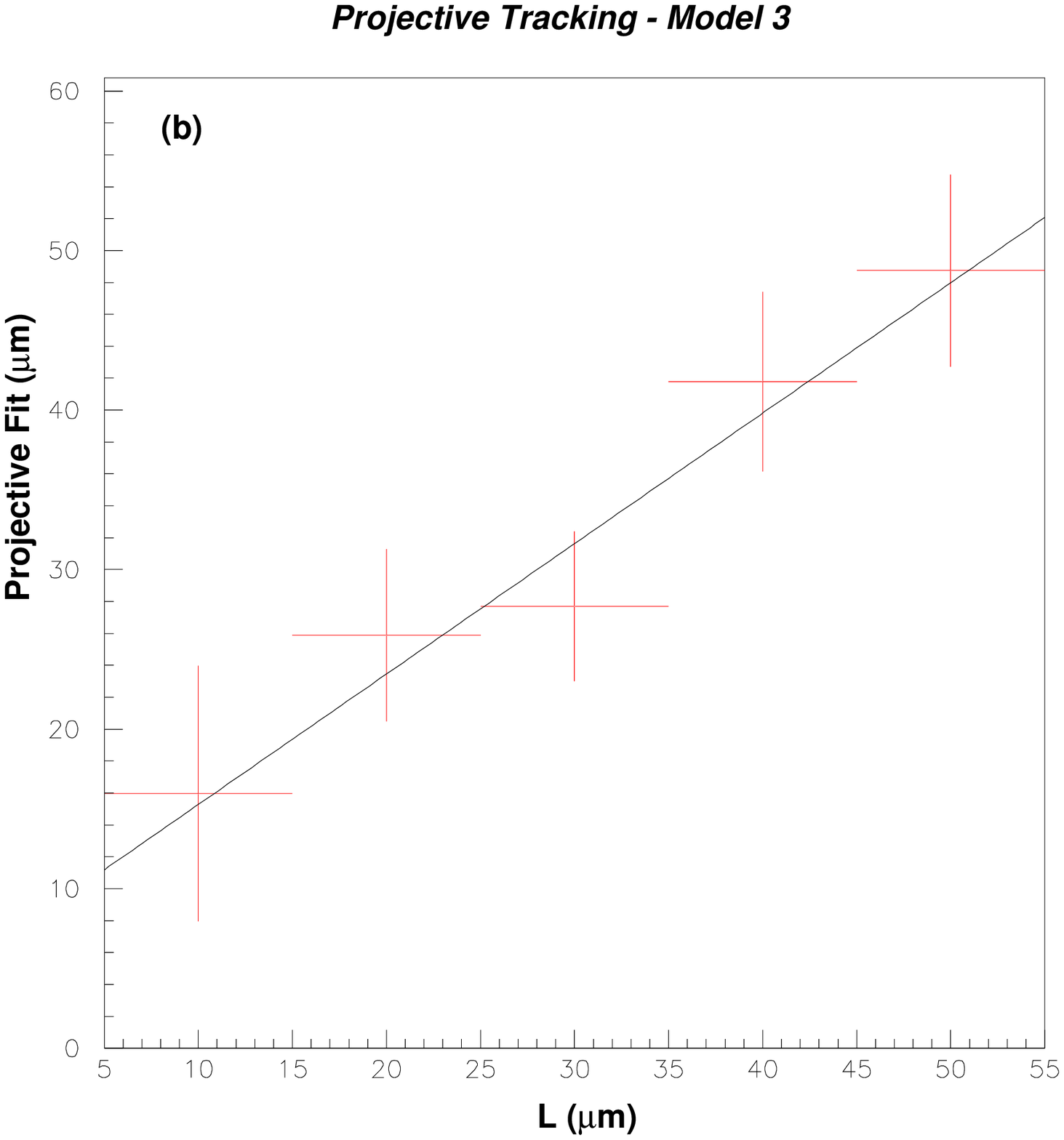} 
}
\caption{\sl (a) Reconstructed projective radial distances, $r$, of the 
$\NI\ra f \bar{f}\G$ decay vertex, for a LC run on a short-lifetime model  
with $L = 10$ $\mu$m together with a fit to an exponential plus constant. 
(b) Shows the results of the corresponding fits for a range of L values.} 
\label{fig:three}
\end{figure}

 For $L$ less than a few cm, we used tracking for measuring 
the vertex of $\NI\ra\G f \bar{f}$ decays. 
When $L$ is very short, less than a few hundred $\mu$m, the beamspot
size becomes important and a 3D procedure is not appropriate.  
Instead, the reconstructed vertex was projected onto the 
$xy$ plane, where the beamspot size is very small, and we 
used the resulting distributions to measure the $\NI$ lifetime.  
We studied several GMSB models with $c\tau_{\NI}$ in the allowed range
and found that the intrinsic resolution of the method was approximately 
10 $\mu$m. An example of the reconstructed 2D decay length distribution 
for a challenging model where the neutralino lifetime can be very 
short~\cite{ourpaper} is shown in Fig.~\ref{fig:three} for statistics 
corresponding to 200 fb$^{-1}$ ($r$ is the $xy$ component of $\lambda$).  

 For 500 $\mu$m $\ltap L \ltap 15$ cm, we used 3D vertexing to determine 
the decay length distribution and hence the lifetime of the $\NI$.  
Vertices arising from $\NI\ra \gamma\G$ and photon conversions in detector 
material were essentially eliminated using cuts on the invariant mass of 
the daughter pairs together with geometrical projection cuts involving the 
mass of the $\NI$ and the topology of the daughter tracks. Methods of 
measuring the $\NI$ mass using the endpoints of photon energies or
threshold techniques, together with details of the projection cuts
have been described~\cite{ourpaper}. Using 200 fb$^{-1}$ of data, 
we concluded that a $c\tau_{\NI}$ measurement with statistical error
of $\sim 4\%$ could be made using this method.

 For $L$ larger than a few cm, we used the $\NI\ra\gamma\G$ channel,
providing much larger statistics. 
The calorimeter was assumed to have pointing capability, using the
shower shapes together with appropriate use of pre-shower detectors.
Assuming the pointing angular resolution mentioned above, we 
demonstrated~\cite{ourpaper} how a decay length measurement can be made. 
We concluded that for lifetimes ranging from approximately 5 cm to 
approximately 2 m this method worked excellently, with statistical precisions 
ranging from a few \% at the shorter end to about 6\% at the upper end of the 
range.
We also investigated the use of timing information to provide a lifetime
measurement, but found it to be less useful than calorimeter pointing.  
However, the use of timing might be relevant to assign purely photonic 
events to bunch crossings and to reject cosmic backgrounds.

 For very long lifetimes, we employed a statistical technique where
the ratio of the number of one photon events in the detector to the 
number of two photon events was determined as a function of $c\tau_{\NI}$.  
This allowed a largely model-independent measurement out to 
$c\tau_{\NI} \simeq$ few 10's m. The possibility of using the ratio of the 
number of no-photon events to one photon events was also 
discussed~\cite{ourpaper}. The latter allows a greater length reach, but 
relies on model-dependent assumptions.

\begin{figure}
\centerline{
\epsfxsize=4in
\epsffile{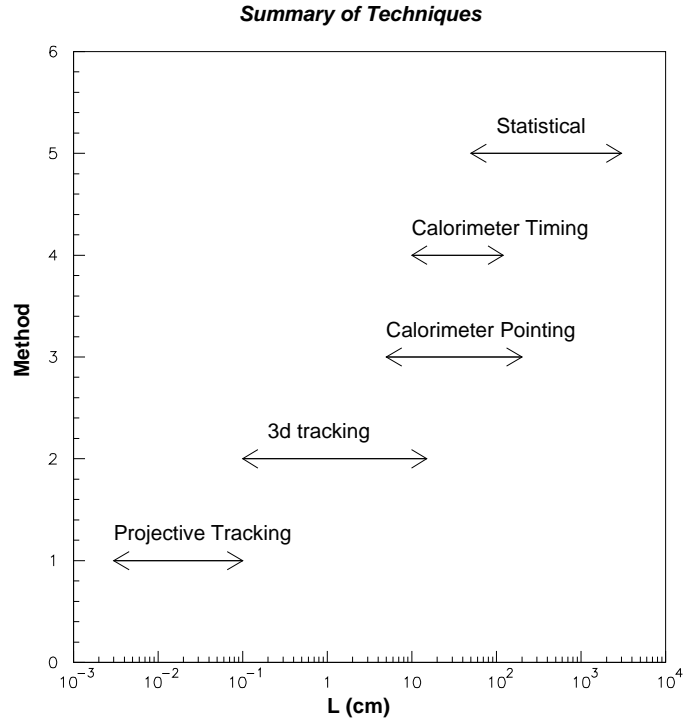} 
}
\caption{\sl Summary of the techniques used here for a $c\tau_{\NI}$ 
measurement at the level of 10\% or better.} 
\label{fig:four}
\end{figure} 

 In Fig.~\ref{fig:four}, we summarise the techniques we have used as 
a function of $L$ for a sample model. The criterion for indicating a 
method as successful is a measurement of $L$ and the $\NI$ lifetime to 
10\% or better. It can be seen that $L$ can be well measured for  
10's of $\mu$m $\ltap L \ltap$ 10's of m, which is in most cases enough 
to cover the wide range allowed by theory and suggested by cosmology. 

 With reference to Eq.~(\ref{eq:NLSPtau}), we note that a 10\% error
in $c\tau_{\NI}$ corresponds to a 3\% error in $\sqrt{F}$.
This is of the same order of magnitude as the uncertainty on the factor 
${\cal B}$, which parameterises mainly the different possible $\NI$ physical 
compositions in GMSB models (cfr. Fig.~\ref{fig:two}a). 
We also checked explicitly that, in comparison, the contributing error from 
a neutralino mass measurement using threshold-scanning techniques or 
end-point methods is negligible~\cite{ourpaper}. 

 Hence we conclude that, for the models considered and under 
conservative assumptions, a determination of $\sqrt{F}$ with a precision 
of approximately 5\% is achievable at a LC by only performing $\NI$ 
lifetime and mass measurements in the context of GMSB with neutralino NLSP. 
Less model dependent and more precise results can be obtained by 
adding information on the $\NI$ physical composition from other observables, 
such as $\NI$ decay BR's, cross sections and distributions.

\end{document}